\documentclass[prb,twocolumn,showpacs,amsmath,amssymb,superscriptaddress]{revtex4}
\usepackage{graphicx}
\begin{document}


\title{Evidence of a photoinduced non-thermal superconducting-to-normal-state phase transition in overdoped Bi$_{2}$Sr$_{2}$Ca$_{0.92}$Y$_{0.08}$Cu$_{2}$O$_{8+\delta}$}

\author{G. Coslovich}

\affiliation{Department of Physics, Universit\`a degli Studi di Trieste, Trieste I-34127, Italy}
\affiliation{Laboratorio Nazionale TASC, AREA Science Park, Basovizza Trieste I-34012, Italy}

\author{C. Giannetti}
\affiliation{Department of Physics, Universit\`a Cattolica del Sacro Cuore, Brescia I-25121, Italy}

\author{F. Cilento}
\affiliation{Department of Physics, Universit\`a degli Studi di Trieste, Trieste I-34127, Italy}
\affiliation{Laboratorio Nazionale TASC, AREA Science Park, Basovizza Trieste I-34012, Italy}

\author{S. Dal Conte}
\affiliation{Department of Physics A. Volta, Universit\`a degli Studi di Pavia, Pavia I-27100, Italy}

\author{G. Ferrini}
\affiliation{Department of Physics, Universit\`a Cattolica del Sacro Cuore, Brescia I-25121, Italy}

\author{P. Galinetto}
\affiliation{Department of Physics A. Volta, Universit\`a degli Studi di Pavia, Pavia I-27100, Italy}

\author{M. Greven}
\affiliation{School of Physics and Astronomy, University of Minnesota, Minneapolis, Minnesota 55455, USA}
\affiliation{Department of Applied Physics, Stanford University, Stanford, California 94305, USA}

\author{H. Eisaki}
\affiliation{Nanoelectronics Research Institute, National Institute of Advanced Industrial Science and Technology, Tsukuba, Ibaraki 305-8568, Japan}

\author{M. Raichle}
\author{R. Liang}
\author{A. Damascelli}
\affiliation{Department of Physics {\rm {\&}} Astronomy, University of British Columbia, Vancouver, British Columbia V6T\,1Z1, Canada}

\author{F. Parmigiani}
\affiliation{Department of Physics, Universit\`a degli Studi di Trieste, Trieste I-34127, Italy}
\affiliation{Sincrotrone Trieste S.C.p.A., Basovizza I-34012, Italy}

\date{\today}

\begin{abstract}

Here we report extensive ultrafast time-resolved reflectivity experiments on overdoped Bi$_{2}$Sr$_{2}$Ca$_{1-x}$Y$_x$Cu$_{2}$O$_{8+\delta}$ single crystals (T$_C$=78 K) aimed to clarify the nature of the superconducting-to-normal-state photoinduced phase transition.  
The experimental data show the lack of the quasiparticles decay time divergence at the fluence required to induce this phase transition, in contrast to the thermally-driven phase transition observed at T$_C$ and at variance with recently reported photoinduced charge-density-wave and spin-density-wave to metal phase transitions.
Our data demonstrate the non-thermal character of the superconducting-to-normal-state photoinduced phase transition. The data have been analyzed using an ad-hoc developed time-dependent Rothwarf-Taylor model, opening the question on the order of this non-equilibrium phase transition.

\end{abstract}

\maketitle

\section{Introduction\label{introduction}}

The possibility of inducing an electronic non-thermal phase transition in high-temperature superconductors (HTSC), by means of ultra-short laser pulses, will set a new path for studying the origin of the superconductivity in these materials.
In facts, under such non-equilibrium conditions, the homogeneous superconducting phase becomes unstable as its free energy increases during the pulse duration\cite{Owen:1972,Carbotte:2003,Giannetti:2009}, while the superconducting order parameter can coexist with the pseudogap or the normal-state.
For many years the exploration of the physics of this process has been a difficult task because of experimental limitations, mainly arising from laser induced heating of the samples\cite{Gay:1999,Gay:PhysC,Gedik:BSSCO,Coslovich:2009}. Only recent all-optical pump-probe experiments on underdoped and optimally doped HTSC\cite{Kusar:2008,Giannetti:2009,Mertelj:2009} have achieved control of the impulsive vaporization of the superconducting condensate in the high-intensity regime. This phenomenon has been observed as the saturation of the transient reflectivity variation ($\Delta $R/R) signal associated with the superconducting phase\cite{Kusar:2008,Giannetti:2009}, in contrast to its linear fluence dependence in the low intensity regime\cite{Kabanov:1999,Dvorsek:2002,Gedik:YBCO,Gedik:BSSCO,Kabanov:2005,Kaindl:2005}. 

A similar photoinduced phase transition (PIPT) to the metallic phase has been recently reported on charge-density-wave\cite{Tomeljak:2009,Yusupov:2010} (CDW) and spin-density wave\cite{Watanabe:2009} (SDW) compounds. While in such systems the PIPT exhibits quasi-thermal character, as it is accompanied by the same critical slowing down observed in quasi-equilibrium conditions at T$_C$\cite{Tomeljak:2009,Watanabe:2009,Yusupov:2010}, the origin of the PIPT in HTSC remains unclear. 
A picture of the photoinduced non-equilibrium state is still lacking due to intrinsic difficulties to disentangle pseudogap and normal-phase signals in optimally and underdoped samples\cite{Giannetti:2009,Mertelj:2009}, the fingerprint of the pseudogap phase being the $\Delta$R/R sign change observed above T$_C$ in pump-probe experiments when probing at 800 nm wavelength\cite{Liu:2008}.

Here we report pump-probe optical reflectivity measurements at 800 nm in the high-excitation regime on an overdoped Bi$_{2}$Sr$_{2}$Ca$_{0.92}$Y$_{0.08}$Cu$_{2}$O$_{8+\delta}$ (Y-Bi2212) single crystal (T$_C$=$78\,$K). At this doping regime the underlying normal phase is Fermi-liquid-like\cite{Lee:rev2006} and no $\Delta$R/R sign change above T$_C$ is measured, at variance with optimally and underdoped samples\cite{Demsar:YBCO99}. These characteristics are fundamental to quantitatively interpret the data within a Rothwarf-Taylor (RT) model with time-dependent parameters.

We show the lack of the quasiparticles decay time increase at the fluence required to photoinduce the transition, at variance with previously reported PIPTs on CDW\cite{Tomeljak:2009,Yusupov:2010} and SDW\cite{Watanabe:2009} systems and in contrast to the decay time divergence observed at T$_C$ on cuprates\cite{Kabanov:1999,Dvorsek:2002,Kabanov:2005}. This finding demonstrates the non-thermal character of the superconducting-to-normal-state PIPT, opening the question on the nature of this phase transition. To address this question we develop a time-dependent RT model. 
The measured decay dynamics is well reproduced by this model, resulting in a non-equilibrium superconducting gap of about one half of the equilibrium value at the pump fluence threshold for the PIPT. 
We propose two different pictures of the non-thermal phase transition occurring in the experiment.

The present work represents a landmark for the growing field of pump-probe techniques, which have been recently extended to the use of several probes, such as Raman scattering\cite{Saichu:2009}, electron-diffraction\cite{Carbone:2008}, angle-resolved photo emission\cite{Perfetti:2007} and broadband optical spectroscopy\cite{Giannetti:Germanato}. All these techniques require an intense ultrashort pump laser pulse, ranging from 0.5 to several mJ/cm$^2$ (Ref. \onlinecite{Perfetti:2007} and \onlinecite{Carbone:2008} respectively), to obtain reliable results.
Our results tackle the long-standing question on the effect of a pump laser pulse at high fluence on the superconducting condensate of HTSC. 

In Sec. \ref{experimental} we briefly report the experimental procedure. In Sec. \ref{low_fluence} we study the temperature dependence of the signal related to the superconducting phase in the low-excitation regime and we show, within the context of the RT model\cite{Rothwarf:1967}, that the decay rate is proportional to the superconducting order parameter $\Delta$. In particular, we focus on the decay time divergence in the vicinity of T$_C$, where $\Delta\rightarrow0$ and thus a vanishing relaxation rate is measured\cite{Kabanov:1999,Dvorsek:2002,Kabanov:2005,Chia:2010}.

In Sec. \ref{discontinuity} we discuss the dynamics as a function of fluence at a fixed temperature (10 K) well below T$_C$. Above a threshold pump fluence, the reflectivity variation deviates from the linear dependence and exhibits a saturation in agreement with previous experiments\cite{Kusar:2008,Giannetti:2009,Mertelj:2009}. This discontinuity is identified with the condensate vaporization in the whole probed volume\cite{Kusar:2008}.
We observe the absence of a decay time divergence above this threshold point at variance with the quasi-thermal case (Sec. \ref{low_fluence}) and with experimental observations of PIPT on CDW\cite{Tomeljak:2009,Yusupov:2010} and SDW\cite{Watanabe:2009} compounds.

To interpret the data in the high-fluence regime we take into account dynamical variations of the non-equilibrium superconducting gap, thus extending the RT model to the time-dependent case (Sec. \ref{RT}).
In Sec. \ref{first-order} we present the experimental data within the analytical results of the RT model. We obtain a linear decrease of the non-equilibrium gap with pump fluence and a finite non-equilibirum superconducting gap 2 ps after the pump pulse that causes the PIPT. 

\section{Experimental methods\label{experimental}}

Pump-probe measurements have been performed on an overdoped Bi$_{2}$Sr$_{2}$Ca$_{0.92}$Y$_{0.08}$Cu$_{2}$O$_{8+\delta}$ single crystal with T$_C$=78 K $\pm$ 5 K. The Y-substituted Bi2212 single crystal was grown in an image furnace by the traveling-solvent floating-zone technique with a non-zero Y content in order to maximize Tc\cite{Eisaki:2004}. The crystal was annealed in flowing oxigen in order to increase the hole concentration and reach the overdoped side of the phase diagram. The sample was subsequently homogenized by further annealing in a sealed quartz ampoule, together with ceramic at the same oxygen content.
In our experiments the 800 nm, 120 fs laser pulses are generated by a cavity dumped Ti:sapphire oscillator. The use of a tunable repetition rate laser source allows us to avoid the experimental problem of average heating effects\cite{Nota:ExpDetails}. These effects prevented earlier observation of the photoinduced condensate vaporization\cite{Gedik:BSSCO,Coslovich:2009}. In the time-resolved experiment the transient reflectivity variation is measured and we denote it as $\Delta$R/R.

\section{Results and Discussion\label{results}}

\subsection{Low-fluence results as a function of temperature\label{low_fluence}}

In the low-intensity regime several common trends have been recognized based on the large amount of experimental data reported on HTSC: i) the appearence below T$_C$ of a $\Delta$R/R signal proportional to the photoexcited quasi-particle (QP) density \cite{Kabanov:1999,Gay:1999,Gay:PhysC,Dvorsek:2002,Gedik:YBCO,Gedik:BSSCO,Kabanov:2005,Kaindl:2005,Kusar:2008,Giannetti:2009,Mertelj:2009,Coslovich:2009,Chia:2010}, ii) an intensity and temperature dependent decay time of this superconducting component \cite{Kabanov:1999,Gay:PhysC,Dvorsek:2002,Gedik:YBCO,Gedik:BSSCO,Kabanov:2005,Kaindl:2005,Mertelj:2009,Chia:2010} iii) the equivalence of the decay time observed probing at 800 nm and the gap dynamics observed in the Thz spectral region\cite{Kaindl:2005}.

\begin{figure} [t]
\includegraphics[keepaspectratio,clip,width=0.5\textwidth] {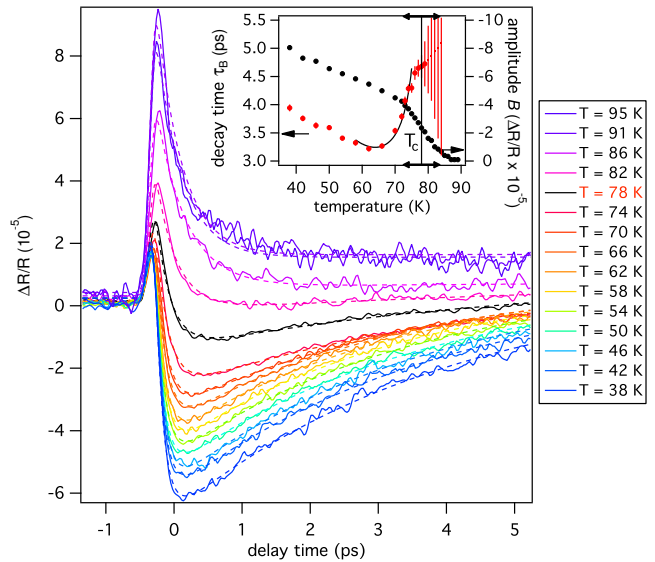}
\caption{\label{vsTscan} $\Delta R/R$ signal (solid lines) as a function of delay time at different temperatures (from 95 K to 38 K) on Y-Bi2212 overdoped single crystal. The dashed lines are the fits to the experimental curves (see the fitting function in the text). The black line represents the $\Delta R/R$ trace at T$_C$. In the inset we report the fit parameters of the SCS, i.e., decay time, $\tau_B$ (red circles), and amplitude, $B$ (black circles). The error bars obtained from the fit procedure are displayed for all the data. The high uncertainty on the decay time above T$_C$ led us to plot just the error bar and not the data point. The black arrow on the temperature axis of the inset shows the superconducting transition amplitude of the sample.}
\end{figure}

In Fig. \ref{vsTscan} we plot the raw ($\Delta$R/R) data of consecutive series of scans taken from 95 K down to 38 K at low pump fluence ($\sim$4 $\mu$J/cm$^2$).
The positive signal above T$_C$ can be reproduced by a single exponential decay function, $Ae^{-t/\tau_A}$, convoluted with the time shape of the laser pulse, describing the relaxation of hot electrons via electron-phonon interaction\cite{Allen:1987,Brorson:2T,Brorson:BISCO} with a relaxation time $\tau_A$ of about 440 fs.
All the traces from room temperature down to 95 K did not show any substantial difference from the 95 K trace. In agreement with the literature\cite{Demsar:YBCO99}, no $\Delta$R/R sign change associated with the pseudogap phase is observed when probing at 800 nm wavelength on the overdoped sample, at variance with optimally and underdoped samples\cite{Liu:2008}. 

The positive and fast signal survives also below T$_C$, superimposed to a negative signal that we recognize as the superconducting signal (SCS) proportional to the photoexcited QP density and that we capture with a single exponential function. The sum of these two functions, $Ae^{-t/\tau_A}+Be^{-t/\tau_B}$, is sufficient to reproduce all the experimental curves as shown in Fig. \ref{vsTscan}. The relevant fit parameters of the SCS, i.e., the amplitude $B$ and the decay time $\tau_B$, are plotted in the inset.

The decay time of the SCS, $\tau_B$, is shown, as a function of temperature, in the inset of Fig.\ref{vsTscan} (red circles).
Starting from 38 K the decay time decreases with temperature, reaching a minimum at 62 K of 3.3 ps. Above this temperature it rapidly increases reaching the value of 4.7 ps at T$_C$. At this temperature the order parameter vanishes ($\Delta\rightarrow0$), the SCS amplitude, $B$, is much smaller than $A$ (the normal-state signal), and the uncertainty on the determination of its decay time strongly increases.
 
The increase in the decay time in the vicinity of T$_C$ is in agreement with previous observations for other HTSC in the low-fluence regime\cite{Kabanov:1999,Dvorsek:2002,Kabanov:2005,Chia:2010} and it has been interpreted as the manifestation of a $\propto1/\Delta$ divergence predicted by several theoretical calculations for BCS superconductors\cite{Tinkham:1972,Schmidt:1975,Kabanov:1999}.
Experimentally, this divergence is smeared out because of the finite superconducting transition amplitude. However, from the analysis of the SCS decay time increase just below T$_C$, we can extract quantitative information about the superconducting gap.

A very useful model to interpret the non-equilibrium dynamics of superconductors in the low-intensity regime is the Rothwarf-Taylor (RT) model\cite{Rothwarf:1967,Kabanov:2005,Demsar:2006} (See Appendix).  In this model, two QPs recombine to form a Cooper pair emitting a boson with energy larger then $2\Delta$. As the reverse process is also allowed, the QP and the boson populations are in quasi-equilibrium and the actual relaxation is determined by inelastic processes.
Within this model, one can write down a set of coupled rate equations, which have analytic solutions in two very important limiting cases, the weak and the strong bottleneck regimes\cite{Rothwarf:1967,Kabanov:2005}. In the first case, the boson inelastic decay rate is fast and the relaxation dynamics is equivalent to simple bi-molecular dynamics\cite{Gedik:YBCO,Kaindl:2005}. In the second regime, namely the strong bottleneck one, the inelastic decay of the boson population strongly slows down the relaxation process. For a given superconductor, the dynamical regime is determined by the particular type of bosons considered in this dynamics.
However, in both regimes and far enough from the critical temperature\cite{nota:Temperatura}, a very simple formula for the QP decay rate $\gamma$ is valid

\begin{equation}
\label{decayRT}
\gamma (T,\Delta(n_{ph}),n_{ph})=(n_{ph}+n_{T})\Gamma(T,\Delta(n_{ph}))
\end{equation}

where $n_{T}$ are the thermal QPs, $n_{ph}$ are the photo-injected ones and $\Gamma(T,\Delta)$ is a function of the microscopic probabilities for the scattering events involving QPs and bosons\cite{Kabanov:2005}.
Given the QP population densities, $n_{ph}$ and $n_{T}$, one can extract from the experimental QP decay rate $\gamma$ the $\Gamma(T,\Delta)$ function in the low-excitation limit. 
We stress that the use of this formula does not imply any assumption on the particular boson involved in the non-equilibrium dynamics.

\begin{figure} [t]
\includegraphics[keepaspectratio,clip,width=0.5\textwidth] {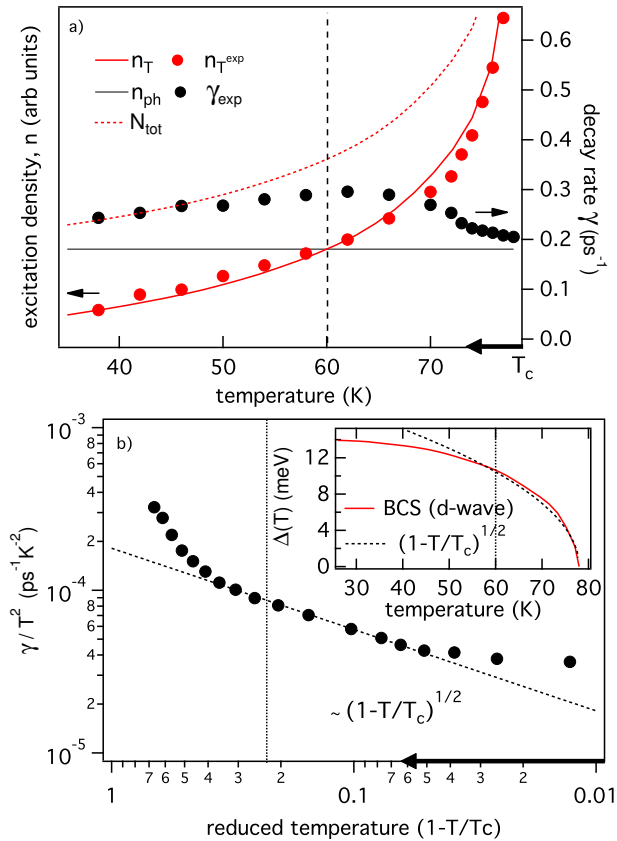}
\caption{\label{VsPhiLow} Panel a) shows the excitation density for both thermal and photoinduced QP, respectively $n_T$ and $n_{ph}$ (solid lines), and their sum $N_{tot}$ (dashed line) as a function of temperature. For the thermal QP we assumed Eq. \ref{Nt} and a d-wave like $\Delta (T)$ dependence and we compared the predicted value with the one obtained  through Eq. \ref{formula16} from the experimental $B(T)$ amplitude (inset of Fig.\ref{vsTscan}). On the right axis we report the experimental decay rate obtained from the fit in Fig. \ref{vsTscan}. 
The b) panel shows $\gamma/T^2$ as a function of the reduced temperature in a double-logarithmic plot. The long-dashed line represents the $(1-T/T_C)^{1/2}$ power-law dependence. 
In the inset we show the result of the numerical integration of the d-wave BCS gap equation as a function of temperature compared to the $(1-T/T_C)^{1/2}$ dependence\cite{Tinkham}. 
In all the panels the vertical short-dashed divide the low from the high temperature regime (see text). In graph b) the error bars are within the black circles size.}
\end{figure}

We use the well-known result obtained by Kabanov et al.\cite{Kabanov:1999} that, for a d-wave superconductor with a 2D Fermi surface with nodes and for temperatures $k_B T<<\Delta\sim5k_B T_C$ (Ref. \onlinecite{Hufner:2008}), the QP population at thermal equilibrium, $n_{T}$, has the form 

\begin{equation}
\label{Nt}
n_{T}=1.64N(0)(k_BT)^2/\Delta
\end{equation}

where N(0) is the density of states at the Fermi level and $\Delta$ is the gap value at equilibrium.
The temperature dependence of $n_{T}$ obtained from Eq. (\ref{Nt}) is plotted in Fig. \ref{VsPhiLow}a. 
To validate Eq. (\ref{Nt}), we estimate $n_T$ from the temperature dependence of the $B$ amplitude (inset of Fig. \ref{vsTscan}) through the formula\cite{Kabanov:2005}

\begin{equation}
\label{formula16}
B (T) \propto \frac{2N_{ph}+n_{ph}}{\sqrt{1+16n_{T}^2+8n_{T}} }
\end{equation}

where $N_{ph}$ is the photoexcited boson population density and we assume, similarly to Refs. \onlinecite{Kabanov:2005,Chia:2010}, the total population density $(n_{ph}+2N_{ph})$ to be constant in temperature since the pump fluence is constant. The $B(T)$ amplitude in the low-temperature limit is measured at T = 10 K.  Good agreement is found between $n_T$ predicted by Eq. (\ref{Nt}) and the value obtained from the experimental data through Eq. (\ref{formula16}) (Fig. \ref{VsPhiLow}a). Thus in the following we will use the value of $n_T$ calculated through Eq. (\ref{Nt}).

In Fig. \ref{VsPhiLow}a the experimental SCS decay rate $\gamma$ and the total QP density, $N_{tot}$=$n_{ph}$+$n_{T}$, are compared. The contribution due to $n_{ph}$ is estimated in the low-temperature limit\cite{Nota:Weight}. 

In the high-temperature limit we observe that: i) the thermal population $n_{T}$ is dominating on $n_{ph}$; ii) $\gamma_{exp}$ is not following the $N_{tot}$ temperature dependence. This finding suggests that the decay time increase observed close to T$_C$ is related to a decrease of $\Gamma(T,\Delta)$. 
We set the separation between the high- and low-temperature regimes at 60 K, i.e, when $n_T>n_{ph}$. Our conclusions are independent of the particular choice of this temperature.

We now verify that the increase in decay time when approaching T$_C$ is related to a real divergence arising from the fact that $\Gamma(T,\Delta)\rightarrow 0$ when $\Delta \rightarrow 0$ and we find the power-law that controls this divergence. 
In Fig. \ref{VsPhiLow}b, we plot the quantity $\gamma_{exp}/T^2$ as a function of the reduced temperature (distance from the critical temperature) on a double-logarithmic scale. Using Eqs. (\ref{decayRT}) and (\ref{Nt}), we find that this quantity is proportional to

\begin{equation}
\label{gamma/T2}
\frac{\gamma}{T^2}\propto \frac{\Gamma(T,\Delta)}{\Delta}
\end{equation}

in the temperature region where $n_T$ is the dominant term in the QP density, i.e., above 60 K.

In this region, we notice a power-law dependence,

\begin{equation}
\label{exponent}
\frac{\Gamma(T,\Delta)}{\Delta}\propto(1-T/T_C)^{\tilde{\beta}}
\end{equation}

with an exponent $\tilde{\beta}$ $\sim$ 1/2 (solid line in Fig. \ref {VsPhiLow}b), which is the same mean-field critical exponent expected for the order parameter $\Delta$ in a BCS superconductor. 
In a d-wave superconductor with T$_C$ of 78 K, the superconducting gap dependence is well approximated by $\Delta\propto(1-T/T_C)^{1/2}$, in the temperature range from 60 to 78 K (see the inset in Fig. \ref {VsPhiLow}b). This assumption is still a good approximation in the case of overdoped HTSC\cite{Lee:2007,Hufner:PRB2008,Yazdani:2009}.
A deviation from this exponent is found a few degrees below T$_C$, since Eq. (\ref{decayRT}) is not applicable in the close vicinity of T$_C$\cite{nota:Temperatura}.

Within the approximate analytic solution (Eq. (\ref{decayRT}))\cite{nota:Temperatura}, we can easily derive the power-law dependence of the $\Gamma(T,\Delta)$ function in the RT approach :

\begin{equation}
\label{gamma}
\Gamma(T,\Delta)\propto(1-T/T_C)^{\alpha} \propto\Delta^{2\alpha}
\end{equation}

where $\alpha$ $\sim$ 1 with an estimated uncertainty of about 30$\%$. This power-law dependence can hardly be derived by first-principles, particularly if the nature of boson involved in the pairing mechanism is unknown.

\subsection{Discontinuity in the fluence dependence\label{discontinuity}}

\begin{figure} [b]
\includegraphics[keepaspectratio,clip,width=0.5\textwidth] {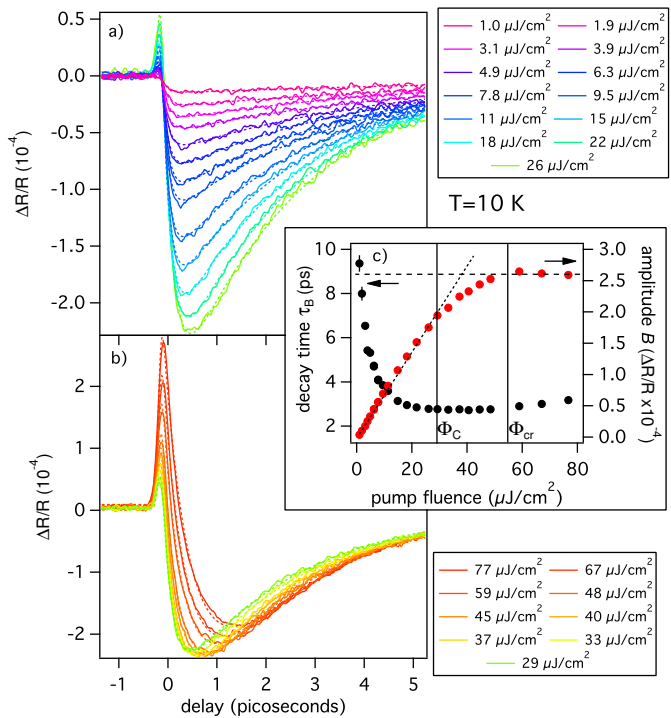}
\caption{\label{vsPhi} $\Delta R/R$ signal, solid lines in panel a) and b), at 10 K at different pump fluences.  The fit to the experimental curves are shown with dashed lines. The fit parameters (decay time $\tau_B$ and amplitude $B$) are reported in panel c) as a function of pump fluence. In panel a) the low-intensity regime is shown, corresponding to $\Phi<\Phi_C$ in panel c). The short-dashed line in panel c) is the linear fit in the low-fluence regime. The panel b) and the panel c) for $\Phi>\Phi_C$ correspond to the high-excitation regime. The long-dashed line in panel c) represent the saturation value. The error bars In panel c) are within the circles size.}
\end{figure}

In Fig. \ref{vsPhi}a and \ref{vsPhi}b, we report the $\Delta$R/R traces at 10 K obtained at increasing pump intensity.
Similarly to Sec. \ref{low_fluence}, all the curves were fitted using two exponential functions convoluted with the time shape of the laser pulse. The results of the fit are superimposed to the experimental curves. Fig. \ref{vsPhi}c shows the SCS amplitude and decay time for each fit as a function of pump fluence.

The low-excitation regime, i.e., the regime where the SCS is proportional to the pump fluence is reported in Fig. \ref{vsPhi}a and in Fig. \ref{vsPhi}c below 26 $\mu$J/cm$^2$ $\equiv$ $\Phi_C$. In agreement with previous works\cite{Kabanov:1999,Gay:1999,Gay:PhysC,Dvorsek:2002,Gedik:YBCO,Gedik:BSSCO,Kabanov:2005,Kaindl:2005,Kusar:2008,Giannetti:2009,Mertelj:2009,Coslovich:2009,Chia:2010}, we assume that the SCS is proportional to the photoinduced QP density, $n_{ph}$. Thus we conclude that $n_{ph}$ increases linearly with the pump fluence in this regime. 
In the zero-fluence limit, where $n_{ph}\rightarrow0$, we observe a divergence of the decay time (Fig. \ref{vsPhi}c). This is related to Eq. (\ref{decayRT}), as the total QP density, ($n_{ph}$+$n_T$), becomes extremely small at low fluence and at low temperature (10 K).

Above $\Phi_C$ the SCS has a sub-linear dependence and we identify this regime as the high-excitation regime\cite{Kusar:2008,Giannetti:2009} (Fig. \ref{vsPhi}b and Fig. \ref{vsPhi}c for $\Phi>\Phi_C$).
In this regime, the SCS exhibits a saturation at a critical fluence, $\Phi_{cr}$, of $\sim$55 $\mu$J/cm$^2$. This saturation means that no more Cooper pairs can be destroyed above $\Phi_{cr}$ and it is considered as the  evidence of the superconducting condensate vaporization during the laser pump pulse\cite{Kusar:2008,Giannetti:2009}.
The fact that the crossover between the linear and the saturated regime does not show an abrupt discontinuity here can be justified by the spatial profiles of the pump and probe light pulses\cite{Kusar:2008}.
The occurence of a real PIPT in this regime has been proved by measuring the emergence of a fast component above $\Phi_{cr}$ in underdoped Bi2212 single crystals\cite{Giannetti:2009}.
Similarly we observe in Fig. \ref{vsPhi}b an enhancement of the positive signal $A$ associated with the fast e-ph free carriers relaxation.

We can compare the SCS in the case of the photoinduced (Fig. \ref{vsPhi}c) and the thermally induced phase transition (inset of Fig. \ref{vsTscan}).
In the former case, the parameter setting the level of perturbation of the system is the pump fluence, while in the latter case this role is played by the sample temperature. Within this analogy, the critical fluence $\Phi_{cr}$, at which the SCS exhibits the saturation, is the counterpart of critical temperature, T$_C$.

According to non-equilibrium superconductivity models\cite{Owen:1972,Parker:1975,Carbotte:2003} a superconducting-to-normal-state PIPT can be either a first-order non-thermal phase transition ($\mu$* model)\cite{Owen:1972,Carbotte:2003} or a quasi-thermal second order phase transition (T* model)\cite{Parker:1975,Carbotte:2003}. 
In the T* model the quasi-thermal condition is applied to electrons and high-frequency bosons, both being distributed with statistics at an effective temperature T*. In both cases the saturation of the SCS reflects the impossibility of exceeding a critical QP density in the stable superconducting phase\cite{Owen:1972,Parker:1975,Carbotte:2003,Giannetti:2009}. 

In analogy to the thermal case (studied in Sec. \ref{low_fluence}), the quasi-thermal photoinduced vaporization would cause the superconducting order parameter to vanish at $\Phi_{cr}$. Thus a diverging decay time should be measured, according to Eq. (\ref{gamma}) and Eq. (\ref{gamma/T2}), and in analogy to experimental results about the PIPT in CDW\cite{Tomeljak:2009,Yusupov:2010} and SDW\cite{Watanabe:2009} compounds.
On the contrary, the experimental results here reported show a decay time that remains finite and below 3.2 ps, thus excluding the quasi-thermal origin of the PIPT and suggesting a finite gap at the threshold fluence.

However, the connection between the decay time of the SCS and the superconducting gap is firmly established only in the low-fluence regime within the analytical results of the the RT model\cite{Kabanov:2005} (Sec. \ref{low_fluence}, Eq. (\ref{gamma})). To extend this concept to the high-fluence regime, when   the non-equilibrium superconducting gap $\Delta(t)$ could strongly vary in time, a validation of the RT model is necessary.

\begin{figure} [t]
\includegraphics[keepaspectratio,clip,width=0.5\textwidth] {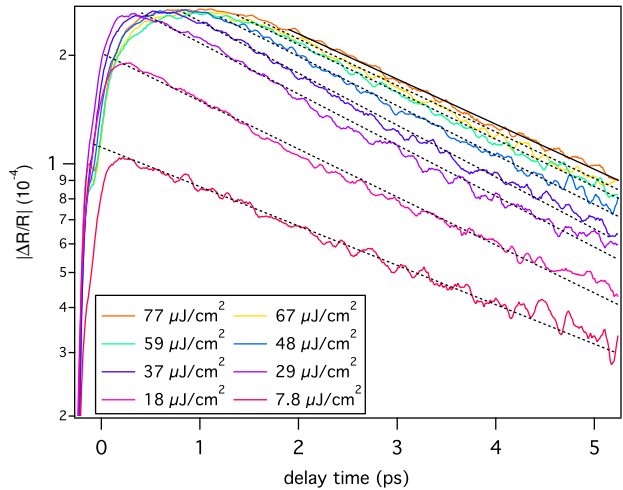}
\caption{\label{ExpDecayRate} We report the $|\Delta R/R|$ traces in logarithmic scale at several pump fluences at T = 10 K, where the exponential $A$ component has been subtracted. The dashed lines represent the exponential fit of the decay. The solid line refers to the curve at $\Phi$ = 77$\mu$J/cm$^2$. }
\end{figure}

Before we discuss this point in Sec. \ref{RT}, we report two important experimental facts about the dynamics of the SCS (See Fig.\ref{ExpDecayRate}),

a) the relaxation dynamics is well reproduced by exponential decays (dashed lines) in the high-fluence regime, as in the low-fluence one. This decay is compatible with the RT model, where the gap is assumed to be constant in time, thus suggesting that the variations of $\Delta(t)$ are small

b) for $\Phi $ $>$ $\Phi_{cr}$, the decay dynamics collapse into a single curve, indicating that the non-equilibrium gap $\Delta(t)$ reaches its minimum value at $\Phi_{cr}$, remaining the same at higher fluences.

\subsection{Rothwarf-Taylor model in the high fluence regime: a time-dependent approach\label{RT}}

When a superconducting system is strongly perturbed through an ultra-short laser pulse, we expect the superconducting order parameter $\Delta$ to have strong variations in time. 
The parameters $\beta$, $\eta$ and $\gamma_{esc}$ of the RT model (see Appendix for definitions) are affected by these variations (Refs. \onlinecite{Kabanov:1999},\onlinecite{Unter:2008}) and can vary in time.
Therefore the high-perturbation limit requires a new time-dependent Rothwarf-Taylor model. Herewith below we use the following assumptions,

i) the time-dependent non-equilibrium superconducting gap $\Delta(n(t))$ can be expressed as a function of $n(t)$ considering the T* and $\mu$* models\cite{Owen:1972,Parker:1975,Carbotte:2003}. In both cases the normalized $\Delta(n(t))$ depends on (1-$an(t)$) (being $n(t)$ the QP density and $a$ a conversion factor) for an s-wave gap symmetry and (1-$an(t)^{3/2}$) for a d-wave gap symmetry.

ii) $\beta(\Delta (n(t)))$ is constant, being its temperature dependence very weak, as reported on YBCO\cite{Gedik:YBCO}

iii) $\eta(\Delta (n(t)))$ is determined with a fit of the rise time at low fluence (See Appendix) and is set to a constant.  
This is a justified by the weak temperature dependence of the pair-breaking time\cite{Kabanov:2005,Kusar:2008} observed in pump-probe experiments at low fluence\cite{Gedik:YBCO,Liu:2008} (see Sec. \ref{low_fluence}) 

iv) $\gamma_{esc} (\Delta (n(t))$ is the only time-dependent parameter and is responsible for the gap-dependence of $\Gamma(T,\Delta)$, evidenced in Sec. \ref{low_fluence} (Eq. \ref{gamma}). The value corresponding to the unpertubed gap, $\gamma_{esc}(0)$, can be determined by a fit of the experimental decay at low fluence using the time-indepedent RT model. (See Appendix)

v) following a well-established trend in the literature\cite{Kabanov:2005,Kusar:2008,Cao:2008,Coslovich:2009} we assume that the cuprates are in the strong-bottleneck regime. We can thus conclude that $\Gamma\propto\gamma_{esc}$ (Ref. \onlinecite{Kabanov:2005}).

Under these approximations, the time-dependent RT equations are:

\begin{eqnarray}
\label{time-depRT1}
\dot{n}=I_{QP}(t)+2\eta p-\beta n^2\\
\dot{p}=I_{ph}(t)-\eta p+\beta n^2/2-\gamma_{esc}(t)\cdot(p-p_T)
\label{time-depRT2}
\end{eqnarray}

with

\begin{equation}
\label{gamma_esc}
\gamma_{esc}(t)=\gamma_{esc}(0)\cdot(\Delta(n(t))/\Delta(0))^{2\alpha}
\end{equation}

where for an s-wave gap symmetry we have:

\begin{equation}
\label{s-wave}
\gamma_{esc}(t)=\gamma_{esc}(0)\cdot(1-an(t))^{2\alpha}
\end{equation}

and for a d-wave gap symmetry:

\begin{equation}
\label{d-wave}
\gamma_{esc}(t)=\gamma_{esc}(0)\cdot(1-an(t)^{3/2})^{2\alpha}
\end{equation} 

with $a$ being a conversion factor which set the perturbation of the non-equilibrium gap by QPs and $\alpha$ the exponent appearing in Eq. \ref{gamma} determined in Sec. \ref{low_fluence} ($\alpha$ $\sim$ 1).
The time-dependent RT equations (Eqs. (\ref{time-depRT1}) and (\ref{time-depRT2})) are then integrated numerically and used to fit the experimental curves of the SCS.

\begin{figure} [t]
\includegraphics[keepaspectratio,clip,width=0.5\textwidth] {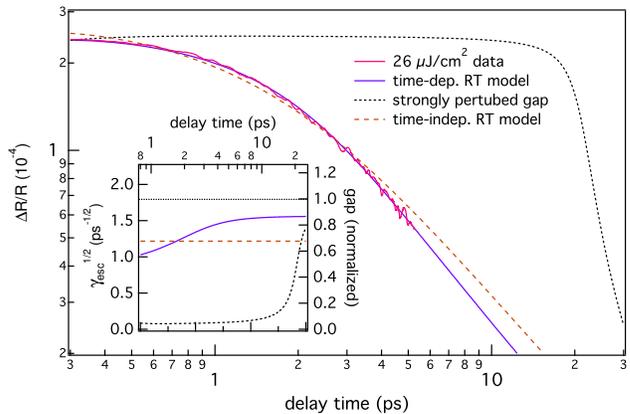}
\caption{\label{confrontoExpRTt-dep} $|\Delta R/R|$ of the SCS at $\Phi$ = 26 $\mu$ J/cm$^2$ and the corresponding fit (solid line) obtained with the numerical solutions of the time-dependent Rothwarf-Taylor equations (Eqs. (\ref{time-depRT1}) and (\ref{time-depRT2})) considering a d-wave gap symmetry. The $a$ optimal value is (1.0$\pm$$0.2$)$\times$10$^{-20}$ cm$^3$. For comparison we also show the fits when the parameter $a$ is fixed to the values $a$ = 2.45$\times$10$^{-20}$ cm$^3$ (short-dashed) and $a$ = 0 (long-dashed), which is equivalent to the time-independent RT model. In the inset we show the corresponding function $\left(\gamma_{esc}(t)\right)^{\frac{1}{2\alpha}}$, which is proportional to the non-equilibrium superconducting gap $\Delta(t)$ (Eq. \ref{gamma_esc}). We normalized the gap to the low-fluence value (See appendix).}
\end{figure}

In Fig. \ref{confrontoExpRTt-dep} we report the results relevant to a single experimental curve at $\Phi$ = $\Phi_{C}$ and T = 10 K, where we assume a d-wave gap symmetry (Eq. (\ref{d-wave})).
The determined $a$ parameter is (1.0$\pm$$0.2$)$\times$10$^{-20}$ cm$^3$. The agreement with the experimental data is very good. The predicted normalized non-equilibrium gap, $\Delta(t)\propto\left(\gamma_{esc}(t)\right)^{\frac{1}{2\alpha}}$, shows a minimum of 60$\%$ at the maximum SCS and an almost complete gap relaxation (85$\%$) after a time delay of 5 ps. Similar results are obtained considering an s-wave gap symmetry (Eq. (\ref{s-wave})).

For comparison, we report also two fits where $a$ is held constant with values $a$ = 0, equivalent to the time-independent RT model, and $a$ = 2.45$\times$10$^{-20}$ cm$^3$, where the gap is suppressed by $\approx$ 95$\%$. In the latter case the predicted relaxation dynamics is extremely slow (of the order of several tens of picoseconds) and strongly non-exponential. Both these characteristics are not compatible with the experimental curve reported in Fig. \ref{confrontoExpRTt}, thus indicating that, below $\Phi_C$, the non-equilibrium gap is never close to zero. 

Above $\Phi_C$, the universal and exponential decay dynamics mentioned in Sec. \ref{discontinuity} suggests a finite non-zero gap after 1-2 ps.
It is possibile to estimate a lower bound for this finite non-equilibrium gap, considering the optimal $a$ parameter and extracting $\Delta(n(t))_{min}$ from the maximum SCS for each curve in Fig. \ref{ExpDecayRate}. The result is that, at any time instant $t$ where the RT model is applicable, the condition $\Delta(t)$ $>$ $\frac{1}{2}\Delta(0)$ is realized.

\subsection{Scenarios for a superconducting-to-normal dynamical phase transition\label{first-order}}

Although the time-dependent RT model discussed in the previous section provides a remarkably good description of the experimental data, we observe that the time-independent RT model is also a valid approximation, provided an average gap value (corresponding to the value predicted by the time-dependent approach at $\sim$ 2 ps delay) is assumed.
In particular we demonstrated that in our experimental conditions we have $\Delta(n_{ph},10K)>\frac{1}{2}\Delta(0,10K)$, thus we can estimate the maximum density of photoexcited QP, $n_{ph}<n_{T=70K}<<\eta/\beta$ (according to the inset of Fig. \ref{VsPhiLow}b and Ref. \onlinecite{Carbotte:2003}). This condition insures the validity of the analytical formulas obtained in Sec. \ref{low_fluence}. 

\begin{figure} [b]

\includegraphics[keepaspectratio,clip,width=0.5\textwidth] {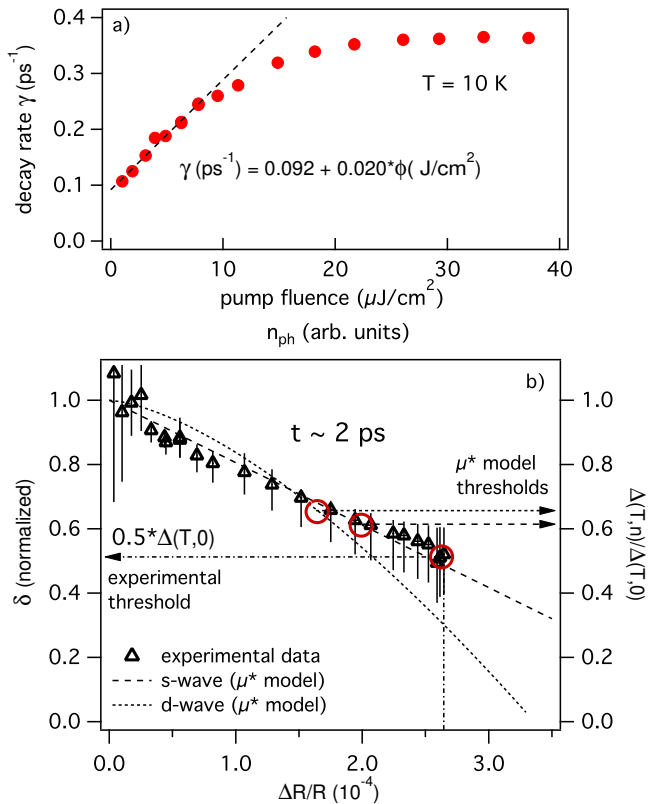}
\caption{\label{VsPhiHigh} a) Experimental decay rate extracted by the fit of the $\Delta$R/R traces at different pump fluences (see Fig. \ref{vsPhi}). The dashed line is the linear fit obtained in the low-intensity limit. The error bars are within the circles size. b) We report $\delta$ as a function of $\Delta R/R$. We normalized the value to the zero fluence limit, which correspond to the equilibrium gap $\Delta (T,0)$. The error bars account both for the 30$\%$ uncertainty in the determination of $\alpha$ and for the difference between the results obtained within the analytic and the numerical solution of the RT equations (see text). On the same graph (right and upper axis) we show the analytical results of the $\mu$* model for the superconducting gap, $\Delta$, as a function of QP density $n$ as reported in Ref. \onlinecite{Carbotte:2003} for both s-wave and d-wave cases. The threshold values are calculated numerically within the $\mu$* model\cite{Carbotte:2003} at a finite temperature corresponding to 10 K in our experiment. }
\end{figure}

We recall Eq. (\ref{decayRT}) and, plotting the decay rate as a function of pump fluence (Fig. \ref{VsPhiHigh}a), we observe the expected linear dependence at low fluence. The intercept of 0.092 ps$^{-1}$ is related to the finite value of $n_T$ at T = 10 K.
This linear dependence on the pump fluence follows the trend observed on underdoped HTSC \cite{Gedik:YBCO,Gedik:BSSCO}, but was not observed clearly on overdoped samples \cite{Gedik:BSSCO,Nota:Slope}. 
We define an effective photoinduced decay rate, $\gamma_{ph}$, 
\begin{eqnarray}
\label{gamma_ph_def}
\gamma_{ph}(T,\Delta(n_{ph}),n_{ph})\equiv\gamma (T,\Delta(n_{ph}),n_{ph})\nonumber\\*
-\gamma_T(T,\Delta (0),n_{ph}=0)
\end{eqnarray}

that represents the decay rate exclusively due to the photoinjected QPs, where the thermal contribution to the decay rate, $\gamma_T$, (intercept in Fig. \ref{VsPhiHigh}a) has been substracted. 
We neglect the sample temperature dependence since in our pump-probe experiment this parameter remains constant. We thus obtain the formula

\begin{widetext}
\begin{equation}
\label{gamma_ph}
\gamma_{ph}(\Delta(n_{ph}),n_{ph})=n_{ph}\Gamma(\Delta(n_{ph}))+n_T[\Gamma(\Delta(n_{ph}))-\Gamma(\Delta(0))]\approx n_{ph}\Gamma(\Delta(n_{ph}))
\end{equation}
\end{widetext}

In the right-hand side of Eq. \ref{gamma_ph} we neglected the second term of the sum. This approximation is valid in the high-excitation regime when $n_{ph}$ is much larger than $n_T$, as long as $\Gamma$ is not going to zero, which is suggested by the absence of a decay time divergence. 

Using Eq. (\ref{gamma}) and the proportionality between $n_{ph}$ and $\Delta R/R$ we can rewrite Eq. (\ref{gamma_ph}) as

\begin{equation}
\label{delta}
\gamma_{ph}=\frac{\Delta R}{R} \Delta^{2\alpha}
\end{equation}

from which we obtain:

\begin{equation}
\label{delta2}
\left(\frac{\gamma_{ph}}{\Delta R/R}\right)^{\frac{1}{2\alpha}}\propto\Delta
\end{equation}

We thus define the quantity $\delta \equiv \left(\frac{\gamma_{ph}}{\Delta R/R}\right)^{\frac{1}{2\alpha}}$, proportional to the non-equilibrium superconducting gap, $\Delta$, and we assume $\alpha$ $\sim$ 1 (Sec. \ref{low_fluence}).

In Fig. \ref{VsPhiHigh}b we report the value of this quantity, normalized to the zero fluence limit value, obtained from the experimental curves of Fig. \ref{vsPhi} as a function of the SCS amplitude. The gap value decreases with $\Delta R/R$, hence it decreases with increasing $n_{ph}$, as expected\cite{Owen:1972,Parker:1975,Carbotte:2003}. On the other side, above the experimental threshold, where $n_{ph}$ is constant, the non-equilibrium gap remains constant, reflecting the correctness of neglecting the T dependence in $\Delta (n_{ph},T)$ and $\gamma_{ph}(T,\Delta,n_{ph})$.

The experimental data are here compared to the predictions of the $\mu$* model, where a first-order phase transition is expected\cite{Owen:1972,Carbotte:2003,Giannetti:2009}.
Within this model the non-equilibrium QP population is described by a Fermi-Dirac distribution with an effective chemical potential $\mu$*, while its temperature remains at the equilibrium value T. In Fig. \ref{VsPhiHigh}b we plot the analytical results for $\Delta(n,T=0)/\Delta(n=0,T=0)$ in the low-excitation limit as a function of the QP density $n$ for both s-wave and d-wave gap symmetry\cite{Carbotte:2003}.
The dependence in the d-wave case is $\propto n^{3/2}$, while it is linear in the s-wave case, as reported in the previous section.

The $\mu$* model predicts a superconducting-to-normal-state phase transition at a finite gap value, when the superconducting-state free energy increases above the normal-state value. At 10 K this phase transition should take place at a relative gap value of $\approx$60$\%$ and $\approx$65$\%$ in the s-wave and d-wave case, respectively\cite{Carbotte:2003} (Fig.\ref{VsPhiHigh}b)

The main features evidenced by the experimental data reported in Fig. \ref{VsPhiHigh} are: i) a finite value of $\Delta(n_{max},T=10\,K)$ of about 1/2 of the equilibrium value at the fluence threshold $\Phi_{cr}$, ii) the linear dependence of $\Delta(n_{ph},T=10\, K)$ with $\Delta R/R$ and thus with $n_{ph}$.

The result i) self-consistently confirms our initial assumption of a non-vanishing $\Gamma$ function and of  $\Delta(n_{ph},10K)$ $>$ $\frac{1}{2}\Delta(0,10K)$. 
The experimental value obtained in this work is remarkably close to the $\mu$* model predictions, even if a quantitative comparison would require a detailed analysis of the photoinduced QP distribution and accurate numerical calculations in the high-fluence regime\cite{Carbotte:2003,Giannetti:2009}.

The result ii) suggests an s-wave like dependence of the non-equilibrium gap. 
Even if this finding seems in contradiction with the equilibrium gap d-wave symmetry reported for HTSC\cite{Hardy:1993}, one should bear in mind that the non-equilibrium population photoinduced by a 100 fs laser pulse in a d-wave superconductor is mostly peaked in the anti-nodal region\cite{Cortes:2010} because of energy-momentum conservation constraints\cite{Howell:2004,Gedik:YBCO}. This leads to a non-thermal effective distribution that can be reproduced by a Fermi-Dirac statistic with anisotropic effective chemical potential.  

The results of such anistropic model are equivalent to those obtained for an s-wave superconductor.
An s-wave like gap symmetry was previously used to explain the temperature dependence of $n_{ph}$ in the low-excitation regime of HTSC\cite{Kabanov:1999} and led to a strong controversy in the pump-probe experiments interpretation\cite{Kabanov:1999,Carbotte:2003}. We remark that this result does not imply an s-wave gap symmetry in HTSC at equilibrium and it is rather related to the excitation process.

Both these results suggest a strongly non-thermal QP population distribution, thus opening two possible physical scenarios behind the superconducting-to-normal PIPT: i) the first-order dynamical phase transition predicted by the $\mu$* model\cite{Owen:1972,Carbotte:2003,Giannetti:2009} and supported by the time-dependent RT model (Sec. \ref{RT}), ii) a second-order transition with a quick ($<$ 1-2 ps) gap recovery, where the fundamental assumptions of the RT model break down on the short timescales in the high-excitation regime. 

The second scenario goes beyond the treatment proposed in Sec. \ref{RT} as the phonon bottleneck would be avoided on the short timescales due to the highly non-thermal QP distribution. 
The break down of the RT equations implies the decoupling of the relaxation dynamics of QP and high-frequency bosons, assessing the non-thermal character of the PIPT, in contrast with the quasi-thermal PIPT predicted by the T* model\cite{Parker:1975,Carbotte:2003}. Within the first picosecond the superconducting state is recovered with a finite non-equilibrium gap, as demonstrated in this section.

However the time-dependent RT model reproduces very precisely the experimental dynamics at $\Phi_C$ (Fig. \ref{confrontoExpRTt-dep}, Sec. \ref{RT}), thus suggesting the validity of this model at excitation levels quite close to the PIPT. Thus, while the non-thermal second-order phase transition implies an abrupt break down of this model between $\Phi_C$  and $\Phi_{cr}$, the non-thermal first-order transition is compatible with the time-dependent RT model also above $\Phi_{cr}$.


The results reported in this manuscript are obtained assuming the validity of Eq. (\ref{decayRT}), which is an approximated analytic solution of the time-independent RT equations\cite{nota:Temperatura}. We repeated the same procedure starting from the numerical solution of the time-indepedent RT equations, which is valid in the whole temperature and fluence range. However the final results of our work are unaffected. The error bars in Fig. \ref{VsPhiHigh}b account for the deviation of the analytic results from the numerical one and for a 30$\%$ uncertainty in the determination of $\alpha$.

Both the hypothesis proposed in this section have a non-thermal character, at variance with the quasi-thermal PIPT reported on other systems, such as CDW\cite{Tomeljak:2009,Yusupov:2010} and SDW\cite{Watanabe:2009}. Since the photoexcitation process is the same, this remarkable difference calls for a deeper understanding of the non-equilibrium QP and bosons thermalization processes in HTSC in respect to CDW and SDW compounds.

\section{Conclusion\label{conclusion}}

We reported pump-probe experiments on an overdoped Y-Bi2212 sample at 800 nm. We explored both the low- and the high-excitation regimes to study the origin of the recently discovered photoinduced vaporization of the superconducting condensate\cite{Kusar:2008,Giannetti:2009}.
We first verified the exponent of the power-law divergence at low-excitation in the vicinity of T$_C$.

In the high-excitation regime, we showed the lack of the quasiparticles decay time increase at the fluence required to photoinduce the transition, in contrast to previously reported PIPTs on CDW\cite{Tomeljak:2009,Yusupov:2010} and SDW\cite{Watanabe:2009} systems.
At that fluence ($\Phi_{cr}$ $\approx$ 55 $\mu$J/cm$^2$) we estimated a finite non-equilibrium gap value (2 ps after the interaction with the pump pulse) of about 1/2 of the equilibrium gap.

This finding proves the non-thermal character of the superconducting-to-normal-state PIPT, leading to two possible phase transition mechanisms: i) a first-order dynamical phase transition predicted by the $\mu$* non-equilibrium superconductivity model\cite{Owen:1972,Carbotte:2003,Giannetti:2009}, and supported by numerical solutions of an extended time-dependent RT model, ii) a second-order transition with a quick gap recovery, where the fundamental assumptions of the RT model abruptly break down in the high-excitation regime on the short timescales (t $<$ 1-2 ps).

These findings tackle the fundamental question on the interaction of an infrared coherent pulse with the superconducting condensate in HTSC at high excitation pump intensity. This is a landmark for the growing field of pump-probe techniques on HTSC\cite{Perfetti:2007,Carbone:2008,Kusar:2008,Giannetti:2009,Mertelj:2009,Saichu:2009,Giannetti:Germanato,Cilento:2010,Chia:2010}. Our experiment defines the maximum pump fluence ($\Phi_{cr}$ $\approx$ 55 $\mu$J/cm$^2$) which still allows to probe the superconducting state of Y-Bi2212; it also demostrates that recent pump-probe experiments perfomed on Bi2212 at higher pump fluences\cite{Perfetti:2007,Carbone:2008,Saichu:2009} are dealing with a dynamical competing admixture of superconducting, normal and possibly pseudogap phases.

The considerable difference for the PIPT in HTSC as compared to CDW\cite{Tomeljak:2009,Yusupov:2010} and SDW\cite{Watanabe:2009} opens an interesting perspective for a deeper understanding of the fundamental distinctions in the non-equilibrium properties of these compounds.

\begin{acknowledgments}
F.C., G.C., and F.P. acknowledge the support of the Italian Ministry of University and Research under Grant Nos. FIRBRBAP045JF2 and FIRB-RBAP06AWK3. The crystal growth work at Stanford University was supported by DOE under Contracts No. DE-FG03-99ER45773 and No. DE-AC03-76SF00515 and by NSF under Grant No. DMR9985067.The work at UBC was supported by the Killam Program (A.D.), the Alfred P. Sloan Foundation (A.D.), the CRC Program (A.D.), NSERC, CFI, CIFAR Quantum Materials, and BCSI.
\end{acknowledgments}

\appendix*
\section{Rothwarf-Taylor equations\label{appendix}}

The non-equilibrium dynamics in superconductors is usually successfully interpreted \cite{Kabanov:2005,Gedik:YBCO} within the phenomenological frame of the Rothwarf-Taylor equations \cite{Rothwarf:1967},
\begin{eqnarray}
\label{RT_appendix1}
\dot{n}=I_{QP}(t)+2\eta p-\beta n^2\\
\dot{p}=I_{ph}(t)-\eta p+\beta n^2/2-\gamma_{esc}(p-p_T)
\label{RT_appendix2}
\end{eqnarray}
describing the density of excitations $n$ coupled to phonons, $p$ being the gap-energy phonon density. The non-equilibrium QP and phonons are photo-injected in the system through the $I_{QP}(t)$ and $I_{ph}(t)$ terms. A gaussian temporal profile of $I_{QP}(t)$ and $I_{ph}(t)$, with the same time-width as the laser pulse, is assumed. The coupling of the electronic and phonon population is obtained through a) the annihilation of a Cooper pair via gap phonon absorption ($p\eta$ term) and b) the emission of gap phonons during the two-body direct recombination of excitations to form a Cooper pair ($\beta n^2$ term). In the phonon bottleneck regime ($\eta$$>$$\gamma_{esc}$) the excitation relaxation is ultimately regulated by the escape rate of the non-equilibrium gap-energy phonons ($\gamma_{esc}(p-p_T)$ term, $p_T$ being the thermal phonon density). The $\gamma_{esc}$ value is determined both by the escape rate of the non-equilibrium phonons  from the probed region and by the energy relaxation through inelastic scattering with the thermal phonons. The $\Delta R/R$ superconducting signal (SCS) is assumed to be proportional to the solution $n(t)$ of Eq.(\ref{RT_appendix1}), in agreement with previous works\cite{Kabanov:1999,Gay:1999,Gay:PhysC,Dvorsek:2002,Gedik:YBCO,Gedik:BSSCO,Kabanov:2005,Kaindl:2005,Kusar:2008,Giannetti:2009,Mertelj:2009,Coslovich:2009,Chia:2010}.
At low fluence ($\Phi<\Phi_C$), the SCS reported in Fig. \ref{ExpDecayRate} is satisfactorily reproduced by considering only the $I_{ph}(t)$ term, i.e. assuming that the pump energy is mainly absorbed through excitation of the phonon population. This result is in agreement with both theoretical predictions within the T* model\cite{Carbotte:2003}  and experimental observations on YBCO\cite{Han:1990,Frenkel:1993}  and MgB$_2$\cite{Demsar:2006}. In the fitting procedure we assume $\beta$ = 0.1cm$^2$/s, as reported in the literature\cite{Gedik:YBCO}. The determined free parameters at low fluence ($\Phi$ = 1 $\mu$J/cm$^2$) are $\gamma$ = (4.5$\pm$$0.5$) ps$^{-1}$ and $\gamma_{esc}$=(3.3$\pm$$0.1$) ps$^{-1}$. These value are compatible with both the results obtained on LSCO \cite{Kusar:2008} and the theoretical estimations of anharmonic processes in YBCO \cite{Kabanov:1999}.


\begin{thebibliography}{51}
\bibitem[{Not()}]{Nota:ExpDetails}
\bibinfo{note}{High pump fluence measurements were perfomed at a repetition
  rate of 108 kHz or less, while low pump fluence ones were obtained at a
  repetition rate of 540 kHz to maximize the signal-to-noise ratio. The total
  photon intensity impinging the sample was never higher than 350 $\mu$W.}
\bibitem[{Not({\natexlab{a}})}]{Nota:Weight}
\bibinfo{note}{To obtain the value of the constant $n_{ph}$ contribution to the
  decay rate we use the experimental values of $\gamma$ at 38 K (Fig.
  \ref{VsPhiLow})a and at 10 K, $\sim$ 4 $\mu$J/cm$^2$ (Fig. \ref{VsPhiHigh}a).
  We assume $\Gamma(T,\Delta)$ to be constant below 40 K (inset of Fig.
  \ref{VsPhiLow}b) and $n_T(10 K)=0$. Thus we have
  $\gamma(38K)=\Gamma_0*(n_T+n_{ph})$ and $\gamma(10K)\sim\Gamma_0*(n_{ph})$,
  from which we derive both $\Gamma_0$ and $n_{ph}$.}
\end{thebibliography}
\end{document}